\documentclass[journal]{IEEEtran}
\usepackage{graphicx}
\usepackage{epstopdf}

\begin{document}

\title{Multi-threaded Geant4 on the Xeon-Phi with Complex High-Energy Physics
       Geometry}

\author{Steven~Farrell,
        Andrea~Dotti,
        Makoto~Asai,
        Paolo~Calafiura,
        Romain~Monnard
        \thanks{Manuscript received November 23, 2015.}
        \thanks{S.~Farrell and P.~Calafiura are with LBL Lawrence Berkeley
                National Laboratory, Berkeley, CA 94720 (USA) (corresponding
                author~-~telephone: 510-486-4181, email: SFarrell@lbl.gov).}%
        \thanks{A.~Dotti and M.~Asai are with SLAC National Accelerator
                Laboratory, Menlo Park, CA 94025 (USA).}%
        \thanks{R.~Monnard is with HES-SO Haute Ecole Sp\'ecialis\'ee de Suisse
                Occidentale, Fribourg (Switzerland).}
        }

\maketitle
\pagestyle{empty}
\thispagestyle{empty}

\begin{abstract}
To study the performance of multi-threaded Geant4 for high-energy physics
experiments, an application has been developed which generalizes and extends
previous work.  A highly-complex detector geometry is used for benchmarking on
an Intel Xeon Phi coprocessor. In addition, an implementation of parallel I/O
based on Intel SCIF and ROOT technologies is incorporated and studied.
\end{abstract}

\section{Introduction}
\IEEEPARstart{I}{n} the midst of the multi-core era, the computing models
employed by high-energy-physics (HEP) experiments must evolve to embrace the
trends of the processor-chip-making industry. As the computing needs of these
experiments---particularly those at the Large Hadron Collider (LHC)---grow,
adoption of many-core architectures and highly-parallel programming models is
essential to prevent degradation in scientific capability.

Simulation of particle interactions is typically a major consumer of CPU
resources in HEP experiments. The recent release of a highly performant
multi-threaded version of Geant4~\cite{Agostinelli2003250}\cite{1610988} opens
the door for experiments to fully take advantage of highly-parallel
technologies.

Intel's many-integrated-core (MIC) processor architecture, known as the Xeon
Phi product line, define a platform for highly-parallel applications. Their
large number of cores and Linux-based environment make them an attractive
compromise between conventional CPUs and general-purpose GPUs. Xeon Phi
processors will be appearing in several next-generation supercomputers such as
Cori at NERSC.

To prepare for these next-generation supercomputers, a Geant4 application
(HepExpMT) has been developed to run multi-threaded HEP particle simulations on
the Xeon Phi. The application serves as a realistic demonstrator of the
capabilities of this advanced architecture for HEP experiments with complex
geometry and parallel writing of particle hit information. It also provides
valuable performance measurements for Geant4 which have already been used to
introduce significant improvements in the memory consumption footprint in
release (10.1)~\cite{g4-v10}.

\section{Many-integrated-core processors}

Intel's MIC product line, the Xeon Phi, is a powerful processor that can
provide good performance for properly optimized applications. The current
generation, known as Knights Corner (KNC), is a coprocessor chip that functions
alongside a traditional CPU and supports both offload and native programming
models. It has more than 50 cores with 512-bit advanced vector instructions
(AVX) running a simplified Linux OS.

The KNC coprocessor doesn't have a hard disk and has a limited amount of RAM
(6-16 GB). A performant communication mechanism between host and coprocessor is
thus essential for applications that produce a significant amount of output.
The Intel Symmetric Communications Interface (SCIF) library serves this
purpose, providing high-performance direct communication and remote memory
access (RMA) operations designed to exploit the full bandwidth capability of
the PCI express bus which connects the host and the coprocessor.

The next generation Xeon Phi, known as Knights Landing (KNL), will provide
significant improvements in compute power and usability. It will be a
standalone fully x86-compatible processor with 72 cores, each one delivering
three times the performance of a KNC core. The KNL also comes with 8-16 GB of
high-bandwidth memory (MCDRAM) and support for up to 384 GB of regular RAM.

Xeon Phi processors already play a significant role in current and future
supercomputers. KNC chips are currently used in machines such as Tianhe-2 at
the National Supercomputer Center in Guangzhou and Stampede at the Texas
Advanced Computing Center (respectively \#1 and \#8 on the current TOP500
list). Supercomputers that are planned to use KNL chips include Cori at NERSC
and Theta at Argonne National Lab.

\section{Multi-threaded Geant4}

Support for multi-threading in Geant4 is available since release version 10.0
(December 2013). The goal of the design is to make efficient use of multi-core
processors and reduce the memory footprint with respect to a sequential
application. The multi-threaded design is based on a master-worker model and
the POSIX standard. Each worker thread is responsible for simulating one or
more full events, thus implementing event-level parallelism. The master thread
is responsible for managing shared data structures and initializing the worker
threads. Threads are independent and require minimal synchronization, which
results in very performant scaling up to the number of physical cores on a
chip.

\section{ROOT I/O} \label{sec:root}

The ROOT~\cite{Brun:1997pa} data analysis framework provides functionality for
writing out HEP event information in a specialized data format (``ROOT
files'').  HEP simulation applications need to write out information describing
particle energy deposits in sensitive detector elements (``hits") in an output
ROOT file.  There are multiple ways to implement output writing when running
events concurrently.
The simplest approach is to have each worker process events independently and
write to separate files on disk which can be merged at the end of processing or
during the analysis stage.  ROOT however provides some functionality for
writing data to a single output file in parallel.  A specialized type of file
called TParallelMergingFile uses sockets to connect clients to a server via TCP
which does the merging of outputs into a single file.

On a KNC coprocessor it is necessary to ship the particle hits output data to
the host to write to disk because of the limited RAM budget on the card. The
TParallelMergingFile implementation can be used for this, but it has been
shown~\cite{RomainThesis} that socket-based communication incurs significant
overhead on the Xeon Phi.  Instead, communication based on the Intel SCIF
library is able to achieve data bandwidth much closer to the theoretical
maximum bandwidth of the PCI express bus~\cite{RomainThesis}.


In order to utilize the high-performance capability of Intel SCIF, a new
backend was written for ROOT. A new ROOT file implementation, TSCIFFile, allows
to use this backend to send data in chunks to the host CPU where a merging
server collects them and merges them to disk~\cite{RomainThesis}.

\section{HepExpMT: an advanced multi-threaded Geant4 benchmark and
         demonstrator}

HepExpMT is an evolution and upgrade of an existing multi-threaded application
(``ParFullCMS''~\cite{Ahn2013,1742-6596-513-2-022005}) developed by Geant4 for
testing code correctness with HEP geometry.
ParFullCMS uses the Xerces-C library~\cite{xercesc} and the Geant4
GDML~\cite{gdml} parser to build a simplified CMS detector geometry provided at
run-time via a GDML file. A uniform magnetic field is applied to the setup and
single particles of a given energy are shot in a random direction and simulated
with the Geant4 physics engine.

The new application has been upgraded for increased complexity and realism, and
also made general enough to run any GDML detector simulation. For performance
studies, the ATLAS detector geometry is used because its very large number of
geometrical elements,
$\mathcal{O}(10^6)$, gives a very challenging setup to test multi-threading
capabilities of Geant4.  The uniform magnetic field along the $z$ axis was not
changed, since for these testing purposed a uniform field is considered
sufficient. The application has been generalized to allow for control of the
primary generation via macro commands to test different aspects of the physics
engine.

The implementation of SCIF-based I/O described in Section~\ref{sec:root} has
been included in HepExpMT to write particle hits output data in parallel. No
sensitive detectors have been implemented in this initial version of the code,
but energy deposits at every particle tracking step are converted into hit data
to write out for each simulated event.  The data is sent regularly to the host
via SCIF RMA where a server process merges the results to a file on local disk.

HepExpMT has been bundled with its support scripts and external libraries in a
standalone package. It is now possible to distribute, compile, and run the
application on different architectures and linux systems without the need of
any external dependency on pre-installed software. The package will be made
public in the near future to allow for users to perform testing of Geant4 and
hardware performance evaluations with custom geometries.


The development and testing of HepExpMT has helped to uncover limitations and
bugs in the underlying software packages, thereby providing valuable feedback
to respective developers. A limitation was identified and patched in the
Xerces-C implementation for extremely large XML files (like the ATLAS GDML).  A
couple of Geant4 bugs related to GDML writing and parsing were fixed.  Finally,
the results of memory consumption measurements prompted significant
improvements to the memory handling in the Geant4 physics code.



\section{Performance measurements}

The performance of the HepExpMT Geant4 application was measured on a 5110p
Knights Corner Xeon Phi with 60 cores and 8 GB of RAM.  Two different GDML
files based on the ATLAS detector were used as input.  The ``full-ATLAS'' GDML
has full detail of the detector except for the ATLAS hadronic end-cap
calorimeter which cannot be represented in GDML form.  A second, simpler
version describing only the ATLAS inner detector (the ``ID-ATLAS'' GDML) is
used for the measurements with output writing.  A uniform field was simulated
at $4~$T, and a fixed number of pions were fired from the interaction point in
random directions with $50~$GeV of momentum.

\begin{figure}
  \centering
  \includegraphics[width=3.3in]{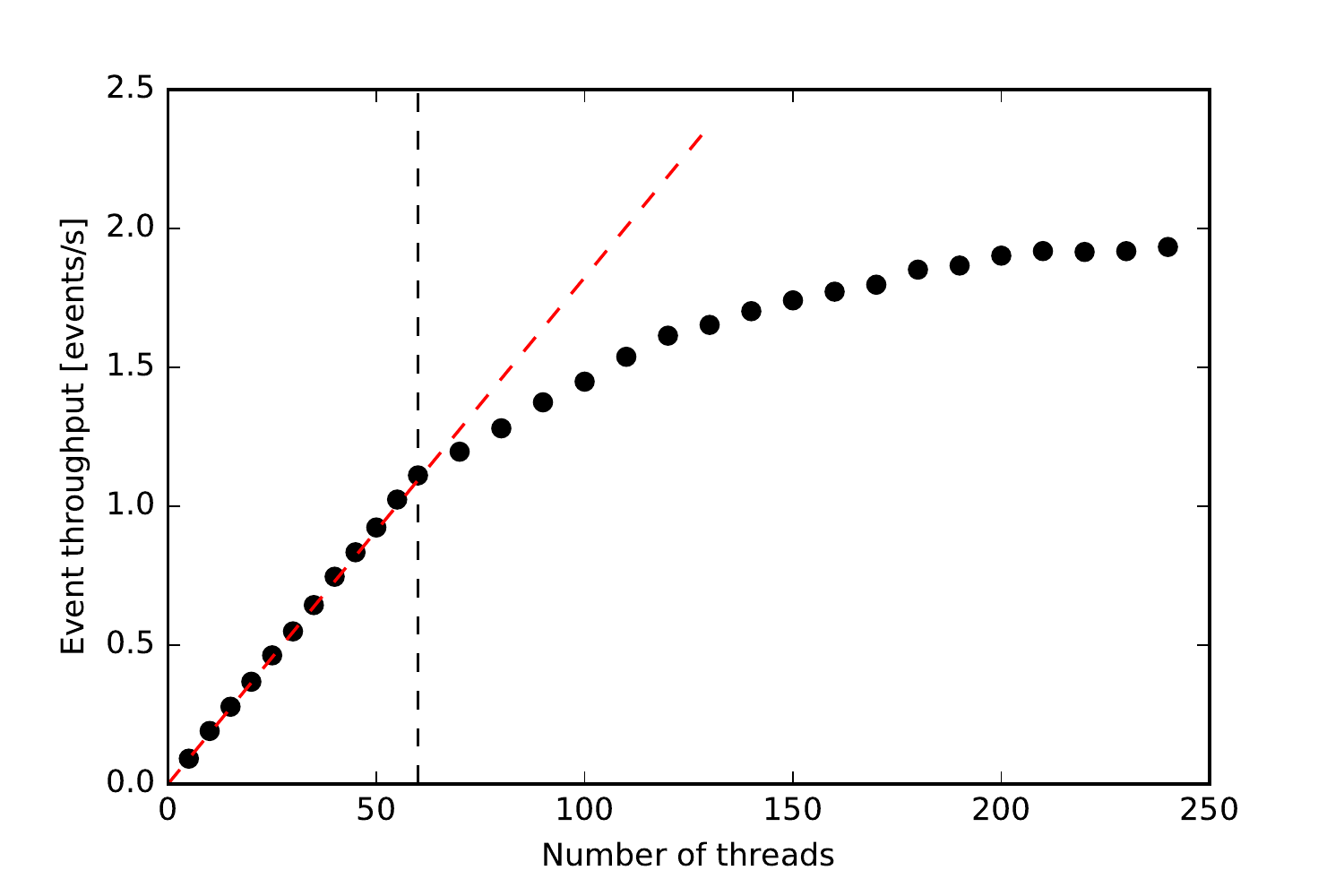}
  \caption{Event processing throughput of the HepExpMT application on the Xeon
           Phi coprocessor as a function of the number of threads. The total
           number of events processed is chosen as 100 times the number of
           threads. Nearly perfect scaling is observed up to the number of
           physics cores on the chip (60, represented by the dashed vertical
           line), with increasing rate up to the maximum 4 threads per core.}
  \label{fig:full_thruput}
\end{figure}

\begin{figure}
  \centering
  \includegraphics[width=3.3in]{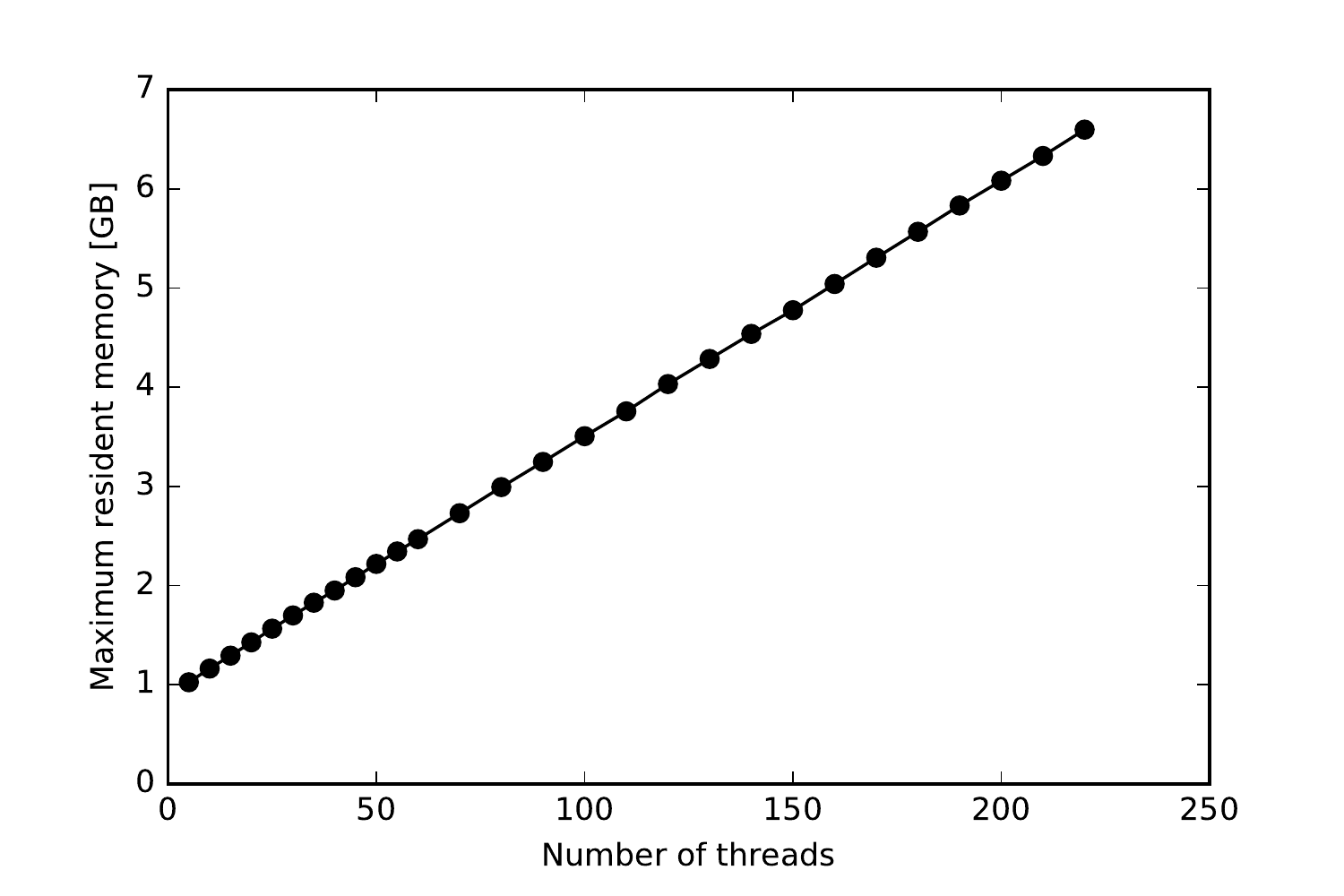}
  \caption{Resident memory of the HepExpMT application on the Xeon Phi
           coprocessor as a function of the number of threads. The number of
           events processed is 100 times the number of threads.}
  \label{fig:full_rss}
\end{figure}

\begin{figure}
  \centering
  \includegraphics[width=3.3in]{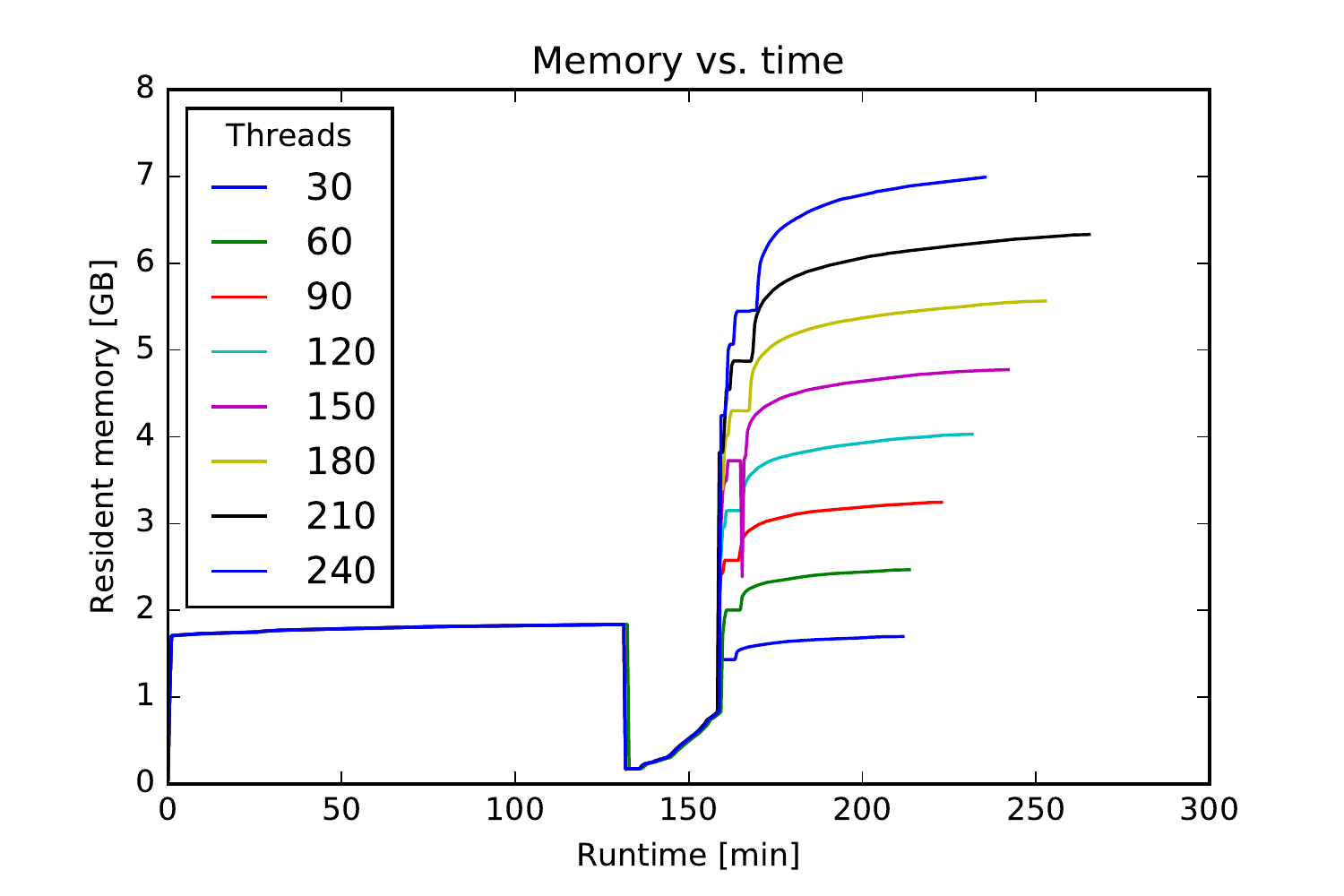}
  \caption{Resident memory of the HepExpMT application on the Xeon Phi
           as a function of application run time for several thread
           configurations. The number of events processed is 100 times the
           number of threads. The long initial plateau is the GDML parsing and
           detector building. The second rise and plateau is the event loop
           processing.}
  \label{fig:full_memTime}
\end{figure}

The first set of measurements use the full-ATLAS detector without I/O to test
the scalability of Geant4 with the most complex detector setup.
Figure~\ref{fig:full_thruput} shows how the event processing rate (throughput)
scales with the number of threads.  The throughput shows nearly perfect scaling
up to the number of cores on the chip (60), showing that the Geant4
multi-threading design is very efficient and introduces minimal overhead and
contention. The throughput continues to increase in the hyper-threading regime
up to the maximum possible 4 threads per core (240 threads).
Figure~\ref{fig:full_rss} shows how the resident memory of the application
scales with the number of threads. The linear increase is expected and shows
that each thread contributes roughly the same amount of memory.  The
coprocessor runs out of memory when running 240 threads, which is an
unfortunate consequence of the limited RAM budget (8 GB) of the KNC
card\footnote{The data-files needed to run the application are copied to the
RAM of the card, no use of NSF has been employed for this study.}.  Finally,
the memory consumption during the runtime of the application is show in
Figure~\ref{fig:full_memTime} for several different threading configurations.
The plot shows a very long plateau from the parsing and processing of the GDML
input, and a rise and plateau during the event loop. This shape is due to the
lazy initialization of memory in the Geant4 physics code.

\begin{figure}
  \centering
  \includegraphics[width=3.3in]{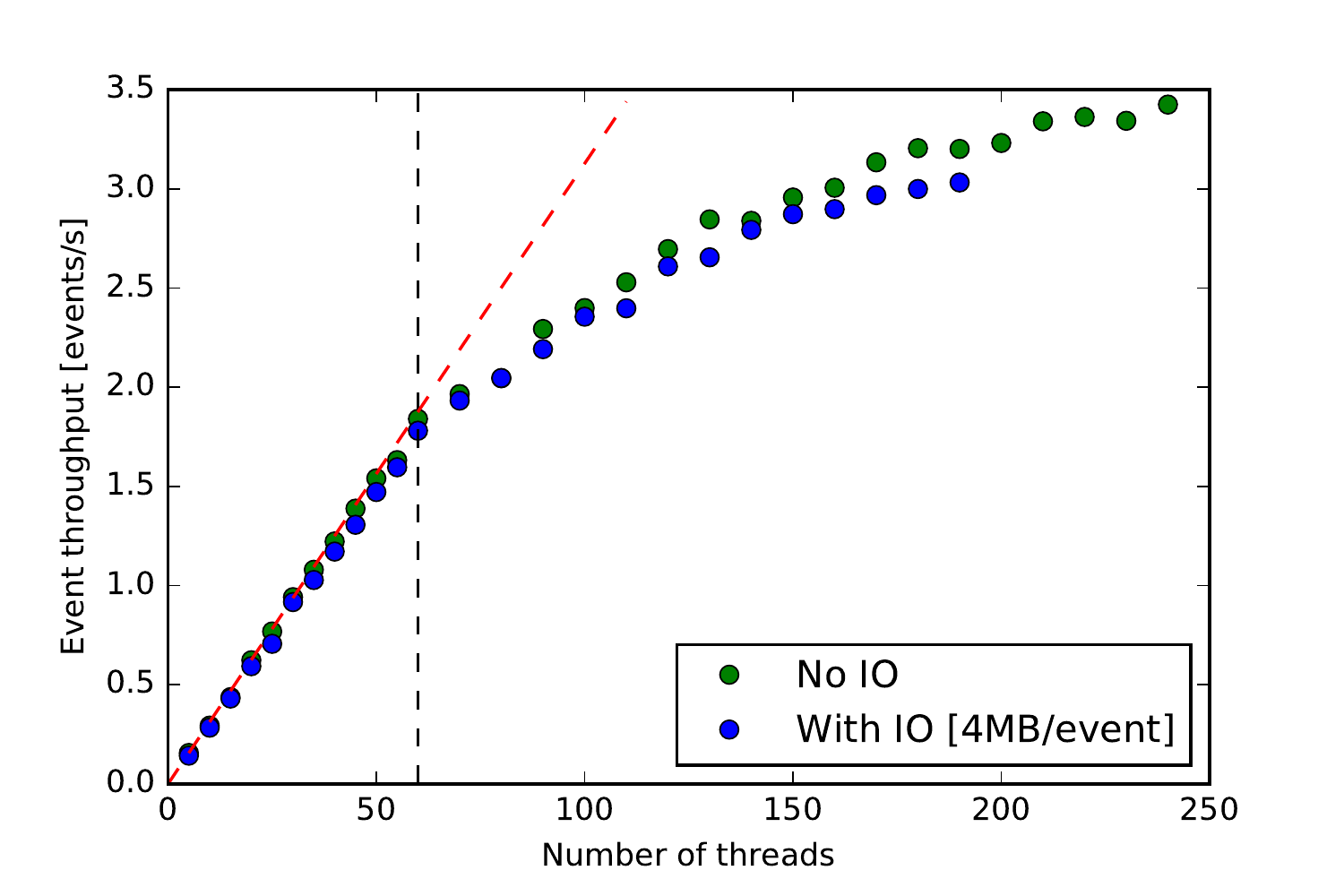}
  \caption{Event processing throughput of the HepExpMT application on the Xeon
           Phi coprocessor as a function of the number of threads, shown with
           and without I/O.  The number of events processed is 50 times the
           number of threads.  No significant impact is observed on throughput
           due to the I/O.}
  \label{fig:id_io_thruput}
\end{figure}

\begin{figure}
  \centering
  \includegraphics[width=3.3in]{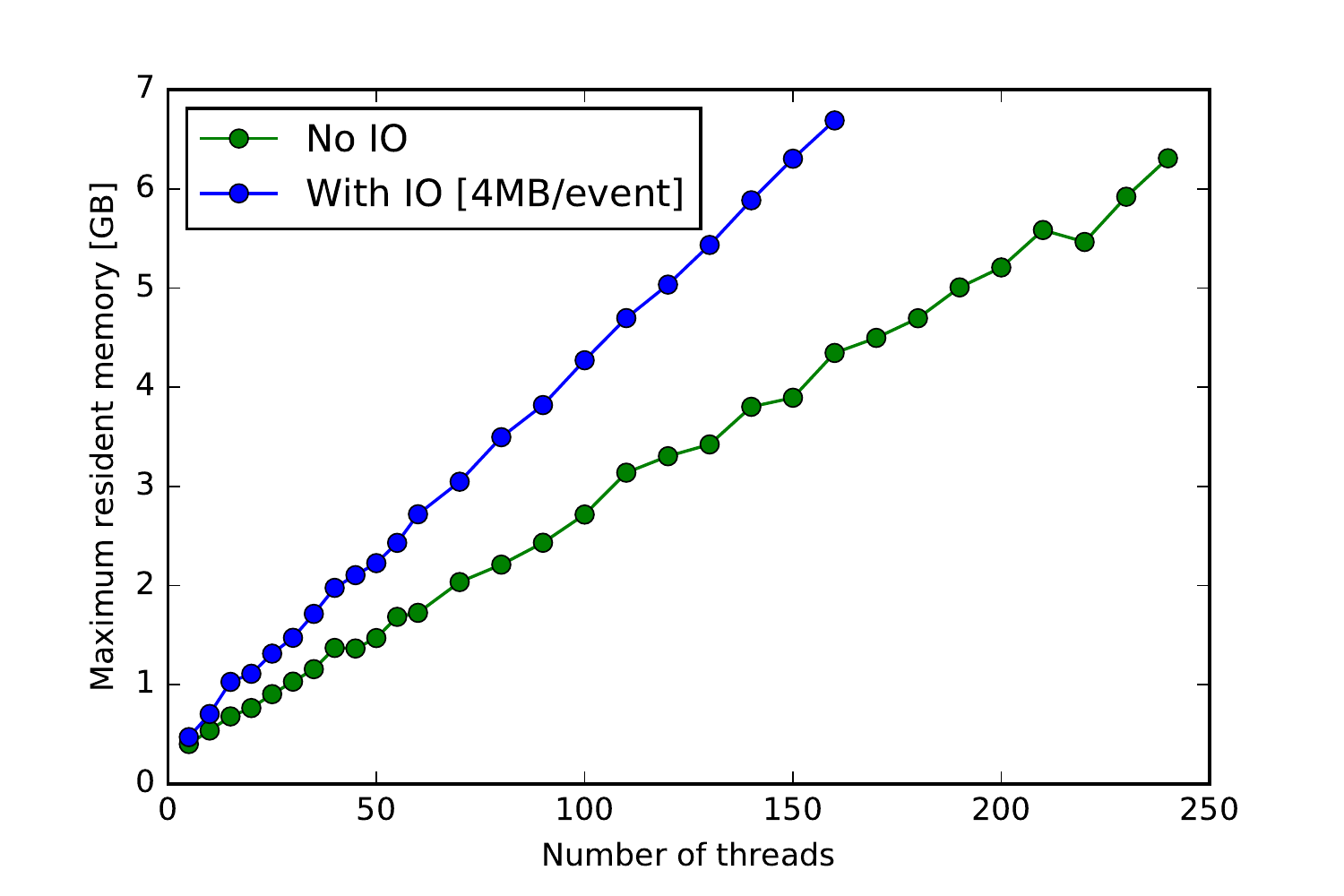}
  \caption{Resident memory of the HepExpMT application on the Xeon Phi
           coprocessor as a function of the number of threads, shown with and
           without I/O.  The number of events processed is 50 times the number
           of threads.}
  \label{fig:id_io_rss}
\end{figure}

\begin{figure}
  \centering
  \includegraphics[width=3.3in]{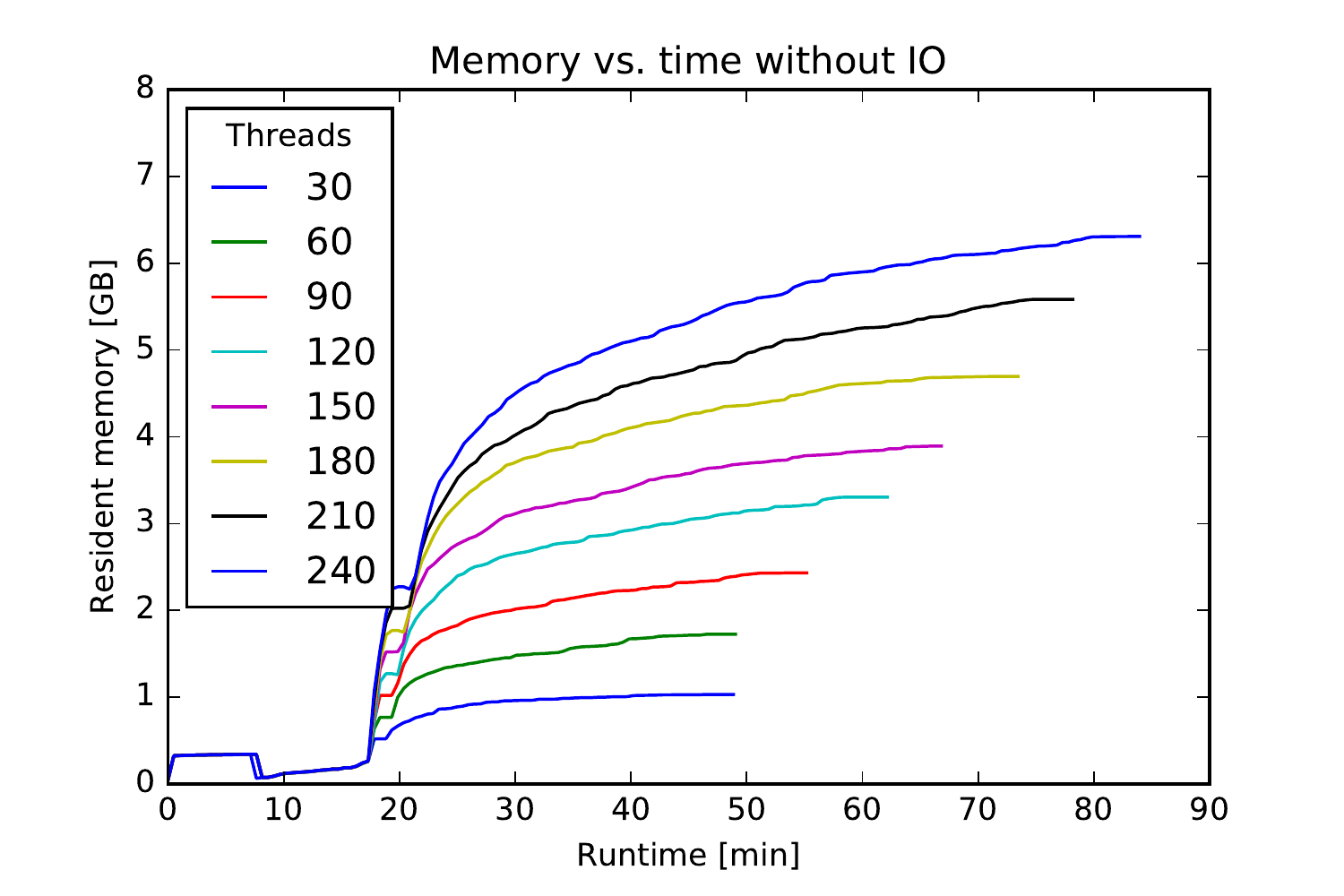}
  \caption{Resident memory of the HepExpMT application with the ID-ATLAS GDML
           and no I/O on the Xeon Phi as a function of application run time for
           several thread configurations and no I/O. The number of events
           processed is 50 times the number of threads.}
  \label{fig:id_noIO_memTime}
\end{figure}

\begin{figure}
  \centering
  \includegraphics[width=3.3in]{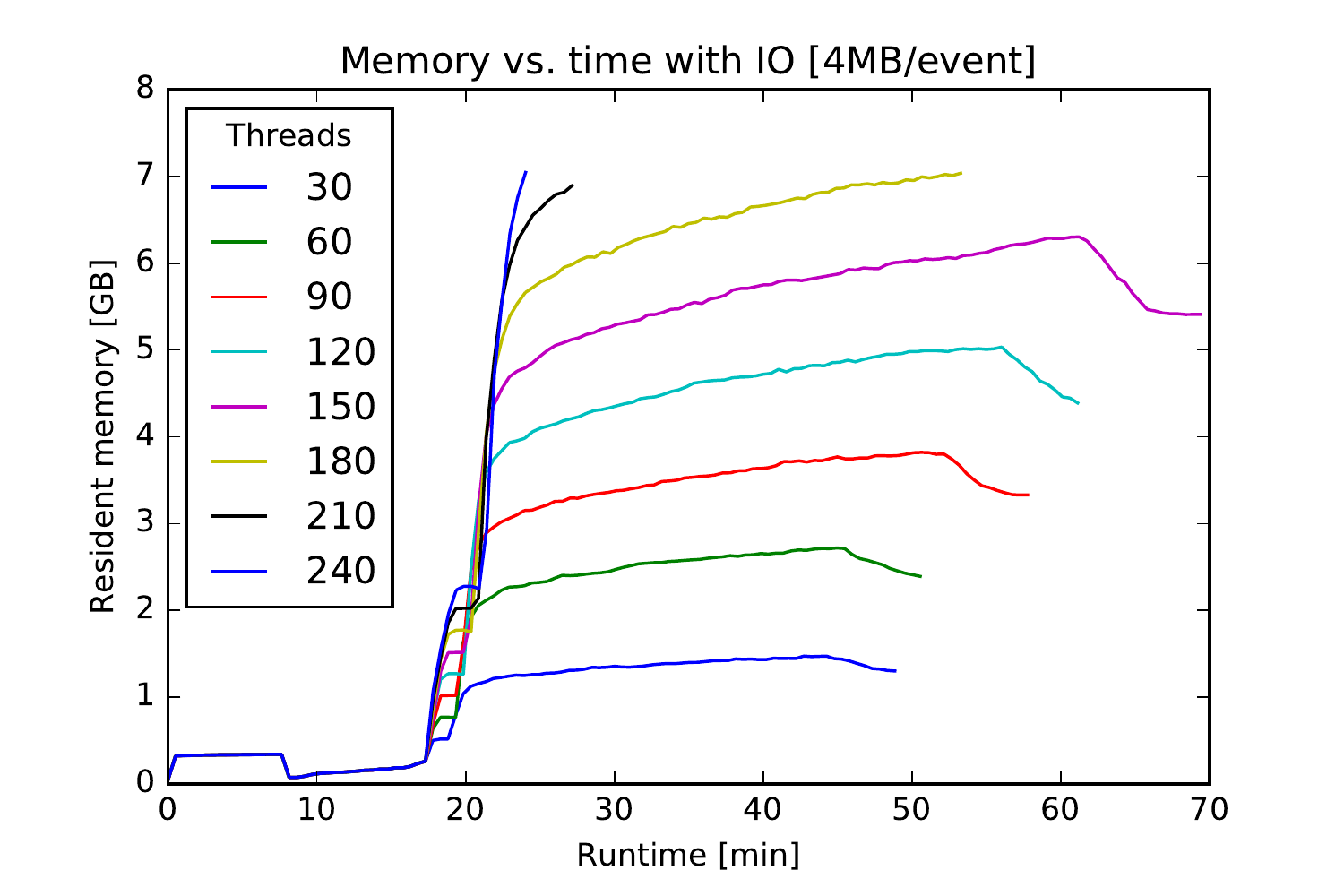}
  \caption{Resident memory of the HepExpMT application with the ID-ATLAS GDML
           with parallel output writing on the Xeon Phi as a function of
           application run time for several thread configurations. The number
           of events processed is 50 times the number of threads.}
  \label{fig:id_io_memTime}
\end{figure}

To measure the impact of parallel I/O on the application performance, the
simpler ID-ATLAS GDML is used. The output data size per event is held fixed at
4 MB when writing with TSCIFFile. Figure~\ref{fig:id_io_thruput} shows the
event throughput comparison with and without I/O.  There is little to no impact
observed from the output writing, which means there are no communication
bottlenecks and no significant overhead in the I/O layer.
Figure~\ref{fig:id_io_rss} shows the linearity of the memory scaling with and
without I/O.  The I/O layer increases the memory consumption by a fixed amount
per thread, which causes jobs to be aborted for fewer number of threads. The
memory consumption during the runtime is shown for this configuration without
I/O in Figure~\ref{fig:id_noIO_memTime} and with I/O in
Figure~\ref{fig:id_io_memTime}. These plots show that the upward shift in
memory consumption of the I/O is fairly flat in time during the event loop.

\section{Conclusions}

This work demonstrates the feasibility of using Xeon Phi for multi-threaded
Geant4 simulations with complex HEP geometry via a new advanced benchmark
application, HepExpMT. Measurements of event throughput and memory consumption
show that Geant4 performs very well with a large number of threads and a
limited memory budget, making it well suited for the MIC architecture. Writing
of event data in parallel using a SCIF-backend for ROOT is shown to perform
well and have no significant impact on event throughput.

This work serves as a valuable learning experience and stepping stone to
prepare HEP experiments for the next-generation Xeon-Phi-based supercomputers
such as Cori. These new machines will be built with the KNL generation of Xeon
Phi, though, which has significant design updates with respect to the KNC
architecture used for these results. In particular, the difficulties with the
tight memory constraints will be relaxed thanks to the increased memory
capacity of the KNL cards. Also, the I/O implementation will likely change
because the KNL cards are self-hosted and can write directly to hard disks or
shared filesystems. However, the parallel I/O mechanisms used in these results
will still be the preferable way to save event data, so the overall scheme may
look similar. New studies will need to be performed when the new Xeon Phi
cards become available.

\vspace{.2in}
\bibliographystyle{IEEEtran}
\bibliography{mybib}{}

\begin{thebibliography}{1}
\providecommand{\url}[1]{#1}
\csname url@samestyle\endcsname
\providecommand{\newblock}{\relax}
\providecommand{\bibinfo}[2]{#2}
\providecommand{\BIBentrySTDinterwordspacing}{\spaceskip=0pt\relax}
\providecommand{\BIBentryALTinterwordstretchfactor}{4}
\providecommand{\BIBentryALTinterwordspacing}{\spaceskip=\fontdimen2\font plus
\BIBentryALTinterwordstretchfactor\fontdimen3\font minus
  \fontdimen4\font\relax}
\providecommand{\BIBforeignlanguage}[2]{{%
\expandafter\ifx\csname l@#1\endcsname\relax
\typeout{** WARNING: IEEEtran.bst: No hyphenation pattern has been}%
\typeout{** loaded for the language `#1'. Using the pattern for}%
\typeout{** the default language instead.}%
\else
\language=\csname l@#1\endcsname
\fi
#2}}
\providecommand{\BIBdecl}{\relax}
\BIBdecl

\bibitem{Agostinelli2003250}
S.~Agostinelli \emph{et~al.}, ``{Geant4}---a simulation toolkit,''
  \emph{Nuclear Instruments and Methods in Physics Research, Section A:
  Accelerators, Spectrometers, Detectors and Associated Equipment}, vol. 506,
  no.~3, pp. 250 -- 303, 2003.

\bibitem{1610988}
J.~Allison \emph{et~al.}, ``{Geant4} developments and applications,''
  \emph{Nuclear Science, IEEE Transactions on}, vol.~53, no.~1, pp. 270--278,
  2006.

\bibitem{g4-v10}
M.~Asai \emph{et~al.}, ``Geant4 version 10 series,'' in \emph{Joint
  International Conference on Mathematics and Computation, Supercomputing in
  Nuclear Applications and the Monte Carlo Methods}, 2015.

\bibitem{Brun:1997pa}
R.~Brun and F.~Rademakers, ``{ROOT: An object oriented data analysis
  framework},'' \emph{Nucl. Instrum. Meth.}, vol. A389, pp. 81--86, 1997.

\bibitem{RomainThesis}
R.~Monnard, ``Concurrent {I/O} from {Xeon Phi} accelerator cards,'' Master's
  thesis, Haute \'{E}cole Sp\'{e}cialis\'{e}e de Suisse Occidentale de
  Fribourg, Switzerland, 2015.

\bibitem{Ahn2013}
S.~Ahn \emph{et~al.}, ``{Geant4-MT: bringing multi-threading into Geant4
  production},'' in \emph{Joint International Conference on Supercomputing in
  Nuclear Applications and Monte Carlo}, vol. 2013, 2013.

\bibitem{1742-6596-513-2-022005}
\BIBentryALTinterwordspacing
G.~Cosmo, ``Geant4 -- towards major release 10,'' \emph{Journal of Physics:
  Conference Series}, vol. 513, no.~2, p. 022005, 2014. [Online]. Available:
  \url{http://stacks.iop.org/1742-6596/513/i=2/a=022005}
\BIBentrySTDinterwordspacing

\bibitem{xercesc}
\BIBentryALTinterwordspacing
 [Online]. Available: \url{http://xerces.apache.org/xerces-c/}
\BIBentrySTDinterwordspacing

\bibitem{gdml}
R.~Chytracek \emph{et~al.}, ``Geometry description markup language for physics
  simulation and analysis applications,'' \emph{IEEE Trans. Nucl. Sci.},
  vol.~53, no.~5, pp. 2892--2896, 2006.

\end{thebibliography}

\end{document}